\documentclass[conference]{IEEEtran}
\IEEEoverridecommandlockouts
\usepackage{cite}
\usepackage{amsmath,amssymb,amsfonts}
\usepackage{algorithmic}
\usepackage{graphicx}
\usepackage{textcomp}
\usepackage{xcolor}
\usepackage{comment}
\usepackage{graphicx}
\usepackage{float} 
\usepackage{subfigure}
\usepackage{multicol}
\usepackage{multirow}
\usepackage{hyperref}
\usepackage[symbol]{footmisc}

\def\BibTeX{{\rm B\kern-.05em{\sc i\kern-.025em b}\kern-.08em
    T\kern-.1667em\lower.7ex\hbox{E}\kern-.125emX}}
\begin{document}

\title{Classify Respiratory Abnormality in Lung Sounds Using STFT and a Fine-Tuned ResNet18 Network}

\author{

\IEEEauthorblockN{Zizhao Chen \textsuperscript{\textsection}}
\IEEEauthorblockA{
\textit{University of Toronto}\\
Toronto, Canada \\
zizhao.chen@mail.utoronto.ca \\
}
\and
\IEEEauthorblockN{Hongliang Wang \textsuperscript{\textsection}}
\IEEEauthorblockA{
\textit{University of Toronto} \\
Toronto, Canada \\
hongliang.wang@mail.utoronto.ca}
\and
\IEEEauthorblockN{Chia-Hui Yeh \textsuperscript{\textsection}}
\IEEEauthorblockA{
\textit{University of Toronto}\\
Toronto, Canada \\
chiahui.yeh@mail.utoronto.ca}
\and
\IEEEauthorblockN{Xilin Liu}
\IEEEauthorblockA{
\textit{University of Toronto}\\
Toronto, Canada \\
xilinliu@ece.utoronto.ca}
}

\maketitle

\begingroup\renewcommand\thefootnote{\textsection}
\footnotetext{These authors contributed equally.}
\endgroup

\begin{abstract}
Recognizing patterns in lung sounds is crucial to detecting and monitoring respiratory diseases. Current techniques for analyzing respiratory sounds demand domain experts and are subject to interpretation. Hence an accurate and automatic respiratory sound classification system is desired. In this work, we took a data-driven approach to classify abnormal lung sounds. We compared the performance using three different feature extraction techniques, which are short-time Fourier transformation (STFT), Mel spectrograms, and Wav2vec, as well as three different classifiers, including pre-trained ResNet18, LightCNN, and Audio Spectrogram Transformer. Our key contributions include the bench-marking of different audio feature extractors and neural network based classifiers, and the implementation of a complete pipeline using STFT and a fine-tuned ResNet18 network. The proposed method achieved Harmonic Scores of 0.89, 0.80, 0.71, 0.36 for tasks 1-1, 1-2, 2-1 and 2-2, respectively on the testing sets in the IEEE BioCAS 2022 Grand Challenge on Respiratory Sound Classification.

\end{abstract}


\section{Introduction}

Respiratory diseases are among the top three global causes of death according to World Health Organization \cite{who2018top}. Studies have shown early diagnosis not only helps prevent the spread of respiratory diseases, but also improves the effectiveness of treatment \cite{pham_phan_king_mertins_mcloughlin_2020}. Clinical studies have identified traits in lung sounds associated with respiratory diseases. Non-invasive, time-saving, and inexpensive medical procedure by auscultation and expert analysis has been developed \cite{Kim2021Respiratory}. However, its broader adaptation is limited by the availability of experienced medical professionals and the subjectivity in the interpretations of lung sound patterns. Thus, there is a need for a consistent and accurate automated respiratory sound classification system.

Recent advances in solving visual and audio classification tasks using neural networks suggest a promising path for data-driven automation design. Machine learning methods have demonstrated the ability to classify lung sounds \cite{Kim2021Respiratory}. In particular, researchers have used conventional machine learning models such as Hidden Markov Models \cite{jakovljevic2017hidden}, Support Vector Machines \cite{SVM}, and Decision Trees \cite{Chambres2018AutomaticDO} to classify lung sounds by first extracting the Mel-frequency cepstral coefficient (MFCC) as features. Many have generated two-dimensional spectrograms, then used them as inputs for different ML architectures such as Convolutional Neural Networks (CNNs) \cite{shuvo2020lightweight, ren2022prototype} and Recurrent Neural Networks \cite{Kochetov2018NoiseMR} to perform the classification task. The top performer \cite{ngo2021deep} on a similar benchmark dataset from the 2017 Internal Conference on Biomedical Health Informatics \cite{rocha2019open} exploited short-time Fourier transformation (STFT) and Gammatone filters, and fed features into an ensemble network of CNN and autoencoder to classify four different respiratory sounds.


In this work, we explored the combination of various feature extraction techniques and classifier architectures. We propose an end-to-end pipeline, R-STFT, for classifying lung sounds, combining STFT and the pre-trained ResNet18 image classifier.

This paper is structured as follows: we first introduce the dataset and the tasks in the IEEE BioCAS 2022 Grand Challenge on Respiratory Sound Classification \cite{ieee_biocas_2022}. Next, we expand on three feature extractors, three neural network classifiers, along with the training procedure. Then we present a comparative analysis of their performance and highlight one combination: R-STFT. Finally, we discuss impacts and identify future improvements.

\begin{figure}[H] 
\centering 
\includegraphics[width=0.5\textwidth]{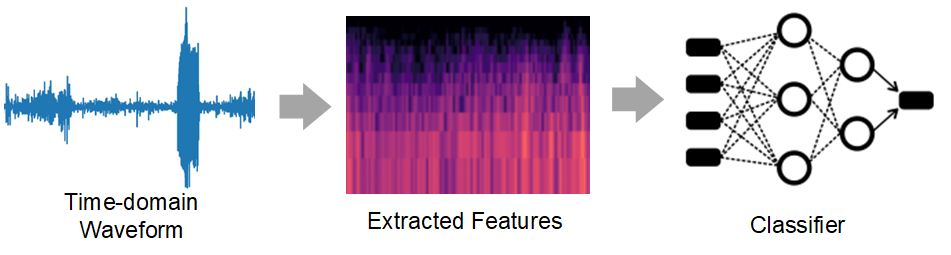}
\caption{ We decompose the task of lung sounds classification into two steps: feature extraction and classification. }
\label{fig:intro}
\end{figure}

\begin{figure*}[!ht] 
\centering 
\includegraphics[width=0.9\textwidth]{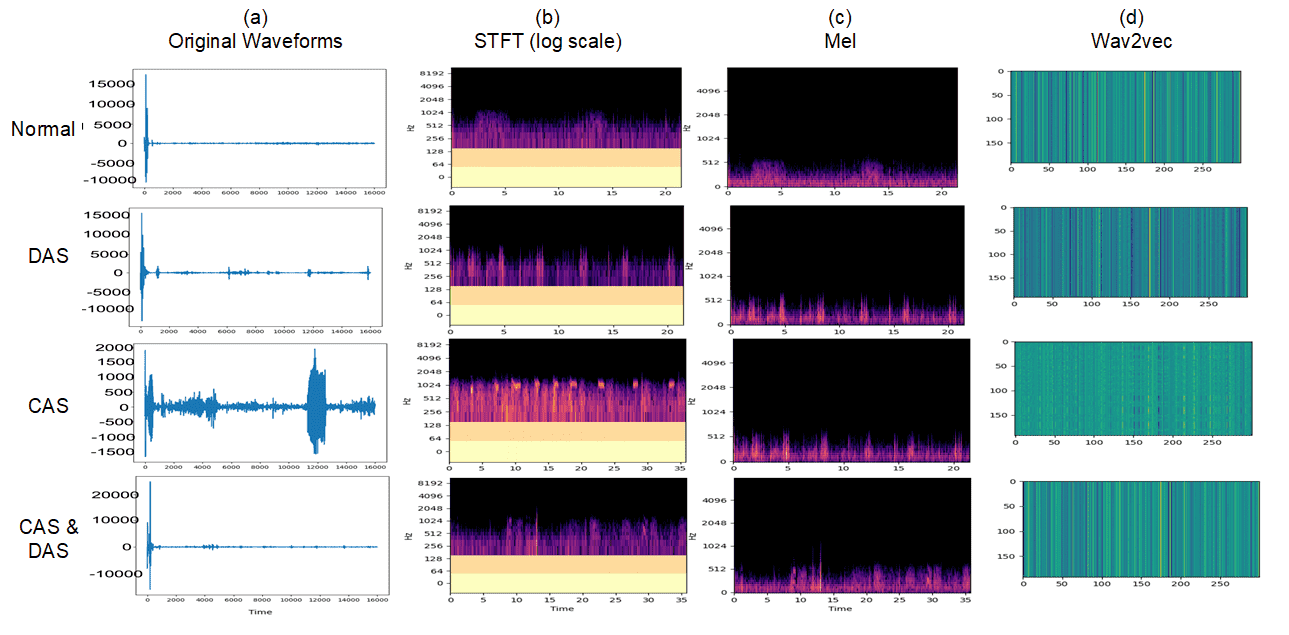}
\caption{(a) Plots of original wavelets. (b) Spectrograms for STFT. (c) Spectrograms for Mel preprocessing. (c) Images of Wav2vec features (no meaningful units). (a), (b), and (c) include plots of Normal, DAS, CAS, C\&D recordings.} 
\label{fig: preprocess}
\end{figure*}

\section{Dataset and Metrics}
The SPRSound: Open-Source SJTU Paediatric Respiratory Sound Database \cite{ieee_biocas_2022} is the first public database that collects data from the Shanghai Children’s Medical Center (SCMC), with ages of children ranging from 1 month to 18 years old. The database contains 6656 labeled events and 1949 labeled recordings. Each recording is segmented into multiple respiratory events, annotated as Normal (N, 77.5\%), Rhonchi (R, 0.6\%), Wheeze (W, 6.8\%), Stridor (S, 0.2\%), Coarse Crackle (CC, 0.7\%), Fine Crackle (FC, 13.7\%), or Wheeze \& Crackle (W\&C, 0.5\%). Recordings are labeled as Normal (N, 66.9\%), Continuous Adventitious Sounds (CAS, 6.9\%), Discontinuous Adventitious Sounds (DAS, 12.7\%), CAS \& DAS (C\&D, 4.4\%) or Poor Quality (PQ, 9.1\%). Recordings are collected at 8kHz for at least 9.2 seconds. This rich dataset provides a unique test ground for evaluating automated lung sound classification systems.

The IEEE BioCAS 2022 Grand challenge on Respiratory Sound Classification proposes four tasks based on this dataset \cite{ieee_biocas_2022}.

\begin{itemize}
    \item Task 1-1: classify events as N or Adventitious.
    \item Task 1-2: classify events as N, R, W, S, CC, FC, or W\&C.
    \item Task 2-1: classify recordings as N or Adventitious.
    \item Task 2-2: classify recordings as N, CAS, DAS, or C\&D.
\end{itemize}

The challenge uses five metrics to evaluate the performance of the classifiers: sensitivity (SE, subscript denotes task), specificity (SP), Average Score (AS), Harmonic Score (HS) and Score. They are defined as follows. 

$$\text{SE}_1 = \frac{\text{R}_\text{r} + \text{W}_\text{w} + \text{CC}_\text{cc} + \text{FC}_\text{fc}+ \text{W\&C}_\text{w\&c}}{\text{R} + \text{W} + \text{CC} + \text{FC} + \text{W\&C}}$$
$$\text{SE}_2 = \frac{\text{CAS}_\text{cas} + \text{DAS}_\text{das} + \text{C\&D}_\text{c\&d}}{\text{CAS} + \text{DAS} + \text{C\&D}}, \ \ \text{SP} = \frac{\text{N}_\text{n}}{\text{N}}$$
$$\text{AS} = \frac{\text{SE} + \text{SP}}{2}, \ \ \text{HS} = \frac{2 \times \text{SE} \times \text{SP}}{(\text{SE} + \text{SP})}, \ \ \text{Score} = \frac{\text{AS} + \text{HS}}{2}$$

The overall pipeline is evaluated by a weighed sum of Scores for each task: $
\text{Total Score} = 0.2 \times \text{Score}_{1-1} + 0.3 \times \text{Score}_{1-2} + 0.2 \times \text{Score}_{2-1} + 0.3 \times \text{Score}_{2-2}
$.

\section{Methods}\label{methods}

We pose the task as a supervised multiclass classification problem. Our solution consists of two components: a pre-processing step that extracts temporal and/or spectral features from the input wave signals, and a neural network-based classifier, as shown in \autoref{fig:intro}.

\subsection{Pre-processing and Feature Extraction}\label{feature-extraction}
Three feature extraction methods were explored in our experiments, which include short-time Fourier transformation (STFT), Mel spectrogram, and Wav2vec (\autoref{fig: preprocess}).

\subsubsection{Short-time Fourier transformation}

STFT is an established technique for extracting frequency features at local sections from temporal signals \cite{liu2021edge}. We selected the Hanning function as the windowing function for Fourier transformation, since adventitious lung sounds are not linear nor stationary. We chose a hop length of 0.01 seconds and a window length of 0.02 seconds between two adjacent Hanning windows, which has been used in a similar lung sound classification task by \cite{ma2019lungbrn}.

\subsubsection{Mel spectrogram}

The Mel scale is a perceptual scale of pitches \cite{stevens1940relation}. Mel spectrograms describe the audio signal in the Mel scale over time. We converted lung sounds to its Mel spectrograms to uncover pitch patterns informative to the domain experts. We used the hop length of 0.01s and a window length of 0.02s, the same configuration as STFT.

\subsubsection{Wav2vec features}

Wav2vec (version 2) learns a speech audio embedded representation by pretraining on 960 hours of audiobooks and fine-tuning its transcripts \cite{DBLP:journals/corr/abs-2006-11477}. It has proven successful in various domains such as speech recognition. We extracted the last layer output of auto speech recognition (ASR) pre-trained models, as the unique features of input audios.



\subsection{Classifier architecture}\label{subsec:classifier}
\subsubsection{LightCNN}
\begin{figure}[!ht]
\centering 
\includegraphics[scale=0.54]{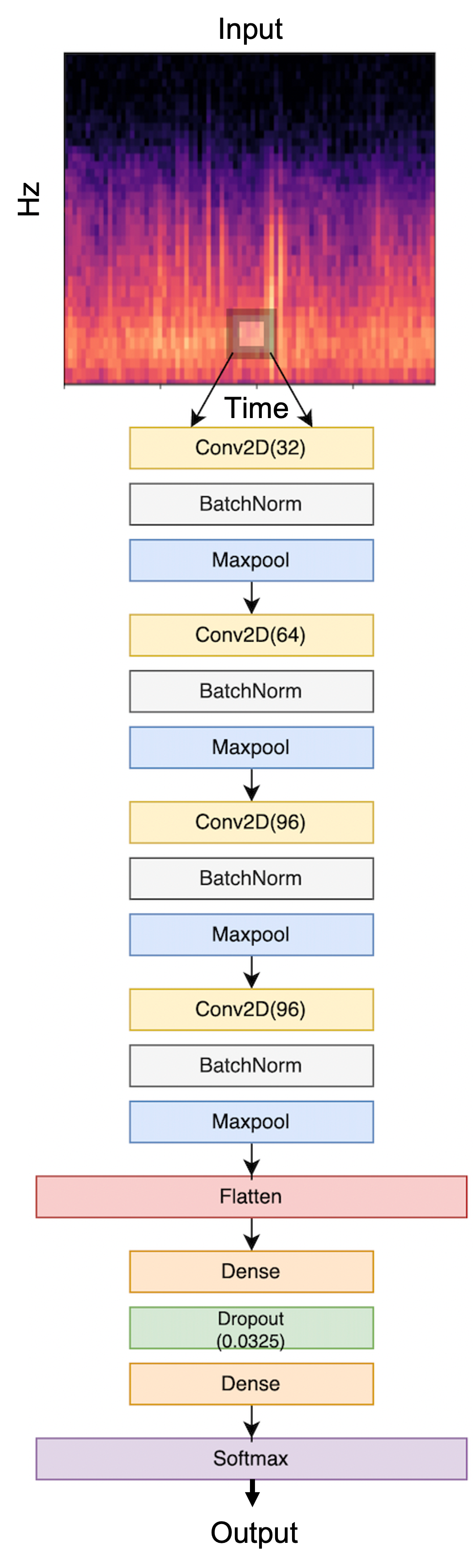} 
\caption{LightCNN model structure diagram} 
\label{fig: LightCNN model structure} 
\end{figure}

We proposed a simple convolutional neural network baseline, inspired by 
the LightCNN model for a similar task \cite{shuvo2020lightweight}. The model structure is illustrated in \autoref{fig: LightCNN model structure}. Briefly, The input layer corresponds to the 3-channel input of 224-by-224 images. The first convolutional layer uses 32 output filters with an 81-pixel square kernel, followed by a 4-pixel max-pooling layer. On top of the first layer, three convolutional layers are stacked, each with a 49-pixel, 25-pixel, 9-pixel kernel containing 64, 96 and 96 channels respectively and corresponding batch-normalization and max-pooling layers with 4-pixel pooling window. Then the CNN-extracted features are flattened in two fully connected layers and linked via a dropout layer (p=0.0325), followed by a SoftMax output layer to finally output the predicted probability for each class.

We chose the default ReLU function as the source of non-linearity \cite{agarap2018deep}. Maximum pooling is applied to reduce feature dimension while retaining spatial invariance \cite{NIPS2012_4824}, \cite{basu2020respiratory}. The batch normalization layers normalize the extracted features, which is a common building block to stabilize neural network training by overcoming internal covariate shift \cite{pmlr-v37-ioffe15}.

\subsubsection{Pre-trained ResNet18}
\begin{figure}[!ht]
\centering 
\includegraphics[scale=0.53]{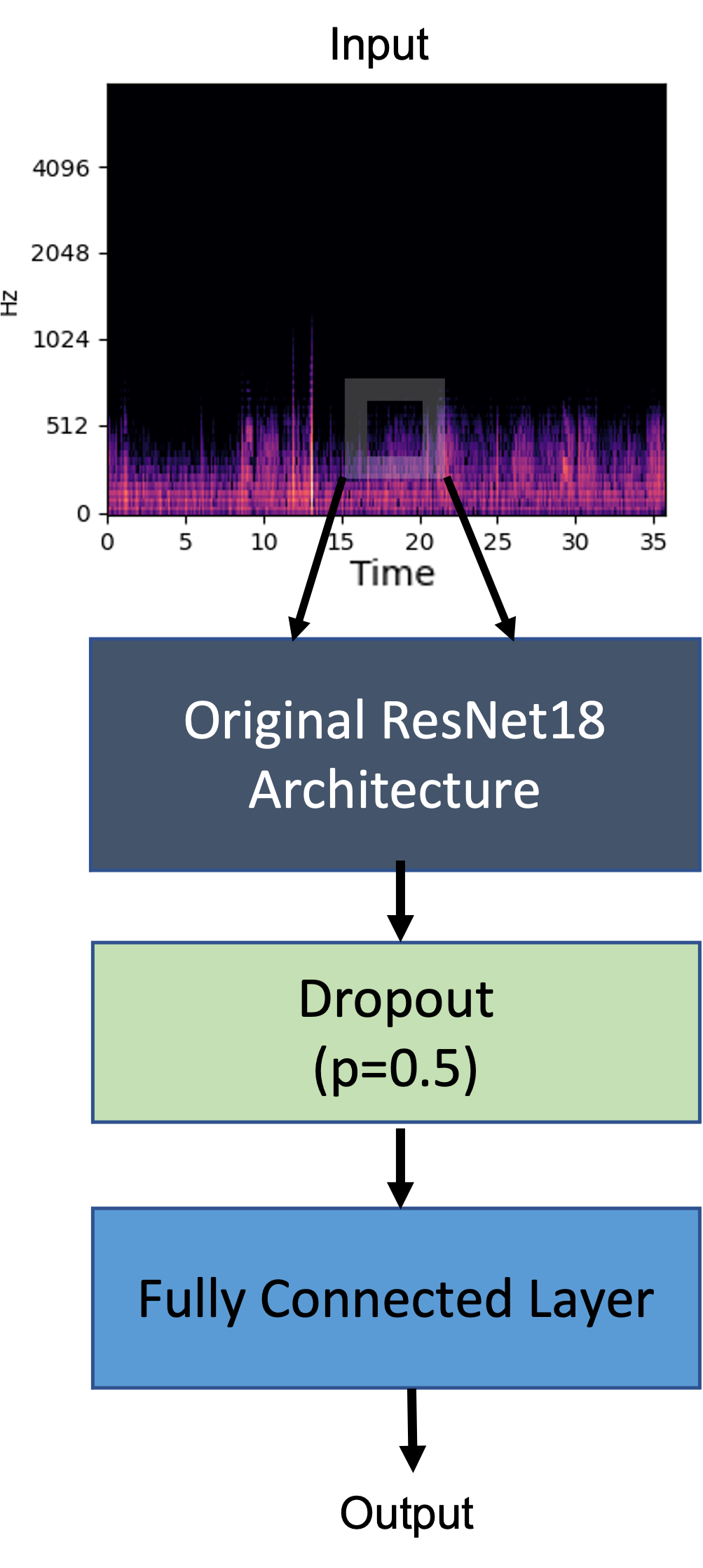} 
\caption{Modified pre-trained ResNet18 model structure diagram} 
\label{fig: Modified pre-trained ResNet18 model structure} 
\end{figure}

Pretraining classifiers on large general datasets and fine-tuning on smaller domain datasets are common practices in the applied machine learning community, thanks to their high training efficiency. Hence, we utilized pre-trained ResNet18 in our experiments. ResNet18 features residual blocks which retain gradients through deep networks. We opted for weights pre-trained on ImageNet1K \cite{deng2009imagenet} provided by the torchvision package in PyTorch ecosystem \cite{paszke2019pytorch}. We added a final dropout layer (p=0.5) and a fully connected layer, on top of the 1000-D outputs of ResNet18. \autoref{fig: Modified pre-trained ResNet18 model structure} illustrates the high-level structure diagram of the modified ResNet18 model.

\subsubsection{Pre-trained Audio Spectrogram Transformer (AST)}
AST has demonstrated its performance on audio classification tasks on AudioSet, a dataset of audio classes in 10 second segments of YouTube videos \cite{gong21b_interspeech}, \cite{45857}. We expect AST to have learned audio-specific features for audio classification than image-based classifiers.

\subsection{Training techniques}\label{subsec:training-techniques}


Weighed loss function was used to balance the over-representation of normal samples in the dataset. Specifically, the weights are proportional to the inverse of the square root of sample size in each target class. Moreover, we dropped the Poor Quality samples in the recording level (Task 2-1 and Task 2-2), as they do not contribute to SE nor SP.

We chose Adam \cite {kingma2014adam} as the optimizer and the learning rate was initialized to 0.001 for CNN models and 0.0001 for pre-trained transformer models, decaying by $10^{-1}$ every 50 epochs. We adapted 9:1 training:validation split. We selected 32 as the batch size unless limited by GPU memory. We ended training or fine-tuning if the validation loss is non-decreasing in 10 epochs. Our experiments were conducted on four NVIDIA® T4 GPUs.


\section{Results}\label{results}
%
\subsection{Task 1: event level classification}
We experimented with all feature extraction techniques and two classifiers for Task 1. The architectures used are the pre-trained ResNet18 and the LightCNN models. In total, we trained four different models, including a pre-trained ResNet18 using STFT as training inputs (R-STFT), a LightCNN using STFT as training inputs (L-STFT), a LightCNN with Mel spectrograms as training inputs (L-MEL), and finally, a LightCNN with Wav2vec as training inputs (L-Wav2vec). As shown in \autoref{tab:task1-results}, R-STFT achieved the best training results for Task 1 (our submission). We have also attached the testing results evaluated by the Challenge committee. The details of the testing results are provided in \cite{ieee_biocas_2022}.


\begin{table}[h]
\centering
\caption{Task 1 results}
\begin{tabular}{|c|c|c|c|c|c|c|c|c|}
    \hline
    \multicolumn{9}{|c|}{Training Results}\\
    \hline
    \multicolumn{1}{|c||}{Model} &\multicolumn{2}{|c|}{R-STFT} &\multicolumn{2}{|c|}{L-STFT} &\multicolumn{2}{|c|}{L-MEL} &\multicolumn{2}{|c|}{L-Wav2vec} \\
    \cline{1-9}
    \multicolumn{1}{|c||}{Task level}& 1-1 & 1-2 & 1-1 & 1-2 & 1-1 & 1-2 & 1-1 & 1-2\\\hline \hline
    \multicolumn{1}{|c||}{SE}& 0.67 & 0.43 & 0.60 & 0.36 & 0.00 & 0.00 & 0.00 & 0.00\\
    \multicolumn{1}{|c||}{SP}& 0.96 & 0.96 & 0.85 & 1.00 & 1.00 & 1.00 & 1.00 & 1.00\\
    \multicolumn{1}{|c||}{AS}& 0.82 & 0.70 & 0.73 & 0.68 & 0.50 & 0.50 & 0.50 & 0.50\\
    \multicolumn{1}{|c||}{HS}& 0.79 & 0.59 & 0.70 & 0.53 & 0.00 & 0.00 & 0.00 & 0.00\\
    \multicolumn{1}{|c||}{Score}& 0.80 & 0.64 & 0.71 & 0.60 & 0.25 & 0.25 & 0.25 & 0.25\\
    \hline\hline
    \multicolumn{9}{|c|}{Testing Results}\\
    \hline
    \multicolumn{1}{|c||}{Model} &\multicolumn{8}{|c|}{R-STFT} \\
    \cline{1-9}
    \multicolumn{1}{|c||}{Task level} 
    &\multicolumn{2}{|c|}{SE} 
    &\multicolumn{2}{|c|}{SP} 
    &\multicolumn{2}{|c|}{AS} 
    &\multicolumn{2}{|c|}{HS} \\
    \hline
    \multicolumn{1}{|c||}{1-1}
    &\multicolumn{2}{|c|}{0.89}
    &\multicolumn{2}{|c|}{0.90}
    &\multicolumn{2}{|c|}{0.89}
    &\multicolumn{2}{|c|}{0.89}\\
    \hline
    \multicolumn{1}{|c||}{1-2}
    &\multicolumn{2}{|c|}{0.68}
    &\multicolumn{2}{|c|}{0.94}
    &\multicolumn{2}{|c|}{0.81}
    &\multicolumn{2}{|c|}{0.79}\\
    \hline
\end{tabular}
\label{tab:task1-results}
\end{table}

\subsection{Task 2: recording level classification}
Similarly, we experimented with the feature extraction techniques and worked with an additional machine learning architecture, the audio spectrogram transformer architecture (AST) since the models that we have experimented in Task 1 (R-STFT and L-STFT) did not achieve comparable results for Task 2. There are four different models trained for this task, which includes R-STFT, L-STFT, a pre-trained ResNet18 with Mel spectrogram as inputs (R-MEL), and an audio spectrogram transformer architecture with Wav2vec feature as inputs (AST-Wav2vec). The training results are presented in Tables \ref{tab:task2-results}.

From Table \ref{tab:task2-results}, R-STFT and L-STFT achieved comparable results for Task 2-1 since they have similar Harmonic Scores. One could argue that R-STFT performed better in Task 2-2 and L-STFT performed better in Task 2-1 since they achieved a higher Score for Task 2-2 and Task 2-1 respectively. However, the team has submitted R-STFT for the challenge since the training results for L-STFT was not yet available at the submission deadline. The testing results of Task 2 for R-STFT is presented in Table \ref{tab:task2-results}.

\begin{table}[h]
\centering
\caption{Task 2 results}
\begin{tabular}{|c|c|c|c|c|c|c|c|c|}
    \hline
    \multicolumn{9}{|c|}{Training Results}\\
    \hline
    \multicolumn{1}{|c||}{Model} &\multicolumn{2}{|c|}{R-STFT} &\multicolumn{2}{|c|}{L-STFT} &\multicolumn{2}{|c|}{R-MEL} &\multicolumn{2}{|c|}{AST-Wav2vec} \\
    \cline{1-9}
    \multicolumn{1}{|c||}{Task level}& 2-1 & 2-2 & 2-1 & 2-2 & 2-1 & 2-2 & 2-1 & 2-2\\ \hline \hline
    \multicolumn{1}{|c||}{SE}& 0.43 & 0.09 & 0.57 & 0.00 & 0.14 & 0.00 & 1.00 & 0.33\\
    \multicolumn{1}{|c||}{SP}& 0.70 & 0.95 & 0.83 & 0.87 & 0.95 & 1.00 & 0.00 & 0.00\\
    \multicolumn{1}{|c||}{AS}& 0.57 & 0.52 & 0.70 & 0.44 & 0.55 & 0.50 & 0.50 & 0.17\\
    \multicolumn{1}{|c||}{HS}& 0.53 & 0.16 & 0.68 & 0.00 & 0.24 & 0.00 & 0.00 & 0.00\\
    \multicolumn{1}{|c||}{Score}& 0.55 & 0.34 & 0.69 & 0.22 & 0.39 & 0.25 & 0.25 & 0.08\\
    \hline\hline
    \multicolumn{9}{|c|}{Testing Results}\\
    \hline
    \multicolumn{1}{|c||}{Model} &\multicolumn{8}{|c|}{R-STFT} \\
    \cline{1-9}
    \multicolumn{1}{|c||}{Task level} 
    &\multicolumn{2}{|c|}{SE} 
    &\multicolumn{2}{|c|}{SP} 
    &\multicolumn{2}{|c|}{AS} 
    &\multicolumn{2}{|c|}{HS} \\
    \hline
    \multicolumn{1}{|c||}{2-1}
    &\multicolumn{2}{|c|}{0.77}
    &\multicolumn{2}{|c|}{0.66}
    &\multicolumn{2}{|c|}{0.72}
    &\multicolumn{2}{|c|}{0.71}\\
    \hline
    \multicolumn{1}{|c||}{2-2}
    &\multicolumn{2}{|c|}{0.23}
    &\multicolumn{2}{|c|}{0.86}
    &\multicolumn{2}{|c|}{0.54}
    &\multicolumn{2}{|c|}{0.36}\\
    \hline
\end{tabular}
\label{tab:task2-results}
\end{table}

\section{Discussion}\label{Sec:discussion}
Overall, R-STFT achieved the best performance for Task 1. For Task 2, the pre-trained ResNet18 models (R-STFT, R-MEL) and the LightCNN model (L-STFT) produced comparable results. We continued to experiment with R-STFT and L-STFT for Task 2, but chose not to train with L-MEL and L-Wav2vec since these two models did not achieve promising results for Task 1. Hence, for Task 2, we tried with different combinations of preprocessing methods and machine learning architectures, which includes R-STFT, L-STFT,  R-MEL and AST-Wav2vec. In general, R-STFT, L-STFT, and R-MEL achieved comparable results for Task 2. On the other hand, AST-Wav2vec did not perform well for Task 2.

R-STFT, L-STFT, and R-MEL achieved a lower sensitivity and a higher specificity score for Task 2. This could be that most examples in the dataset were labeled as Normal, which leads to an imbalanced training set despite our effort to counteract such imbalance with weighed loss function. To improve the SE and SP scores of our classifier, we can consider utilizing a over-sampling technique to increase the samples of other classes \cite{afzal_schuemie_van_blijderveen_sen_sturkenboom_kors_2013}.


\section{Conclusion}

We experimented with different combinations of machine learning models and feature extraction techniques to perform the classification of respiratory sounds. R-STFT yielded a better performance consistently compared to the other combinations, achieving Scores of 0.89, 0.80, 0.71, 0.36 for tasks 1-1, 1-2, 2-1 and 2-2, respectively on the testing sets of the IEEE BioCAS 2022 Grand Challenge on Respiratory Sound Classification. In future work, addressing the imbalanced samples is essential to improving sensitivity scores. Experimenting with more combinations of architectures and feature extraction techniques that better suit the nature of lung sounds, as well as adopting a methodical approach such as hyperparameter sweeping to select the best hyperparameters are both promising directions to further improve our classifier's performance.

\bibliographystyle{IEEEtran}
\bibliography{ref}

\end{document}